# Correlation-aware Coarse-to-fine MLPs for Deformable Medical Image Registration

Mingyuan Meng[1,2], Dagan Feng[1], Lei Bi[2], Jinman Kim[1]
[1] School of Computer Science, The University of Sydney, Australia.
[2] Institute of Translational Medicine, Shanghai Jiao Tong University, China.
mmen2292@uni.sydney.edu.au, {dagan.feng,jinman.kim}@sydney.edu.au, lei.bi@sjtu.edu.cn

## Abstract

*Deformable image registration is a fundamental step for medical image analysis. Recently, transformers have been used for registration and outperformed Convolutional Neural Networks (CNNs). Transformers can capture long-range dependence among image features, which have been shown beneficial for registration. However, due to the high computation/memory loads of self-attention, transformers are typically used at downsampled feature resolutions and cannot capture fine-grained long-range dependence at the full image resolution. This limits deformable registration as it necessitates precise dense correspondence between each image pixel. Multi-layer Perceptrons (MLPs) without self-attention are efficient in computation/memory usage, enabling the feasibility of capturing fine-grained long-range dependence at full resolution. Nevertheless, MLPs have not been extensively explored for image registration and are lacking the consideration of inductive bias crucial for medical registration tasks. In this study, we propose the first correlation-aware MLP-based registration network (CorrMLP) for deformable medical image registration. Our CorrMLP introduces a correlation-aware multi-window MLP block in a novel coarse-to-fine registration architecture, which captures fine-grained multi-range dependence to perform correlation-aware coarse-to-fine registration. Extensive experiments with seven public medical datasets show that our CorrMLP outperforms state-of-the-art deformable registration methods.*

## 1. Introduction

Medical image registration is a fundamental requirement for medical image analysis and has been an active research focus for decades [1, 2]. It spatially aligns medical images acquired from different patients, times, or scanners, which serves as a crucial step for various clinical tasks, such as tumor growth monitoring and group analysis [3]. Due to pathological changes or anatomy variations among patients, medical images carry many non-linear local deformations, especially for complex organs such as the brain's cerebral cortex [4]. Therefore, different from the common natural image registration tasks (e.g., panorama stitching [5]) that aim to remove the global misalignments caused by parallax, medical image registration heavily relies on deformable registration, and this motivates the current research focus [2, 6]. For example, many medical image registration studies assume that the images can be globally aligned after affine registration and mainly focus on deformable registration with non-linear local deformations [7-16].

Deformable image registration aims to find a dense non-linear spatial transformation between a pair of images so that the two images can be spatially aligned with each other. Traditional image registration methods usually formulate deformable registration as a time-consuming iterative optimization problem [17, 18]. Recently, deep registration methods based on Convolutional Neural Networks (CNNs) or transformers have been widely used to perform fast end-to-end registration [3, 6]. These methods learn a mapping from image pairs to spatial transformations based on a set of training data, which have shown superior registration performance than traditional methods [7-16].

The Visual transformer (ViT) [19] and its window-based variant, Swin transformer [20], have been widely adopted in various vision tasks for their great capability to capture long-range dependence via self-attention. This capability has been shown beneficial for deformable registration as it can enlarge the receptive field to model large deformations between images [12-15]. Nevertheless, due to the high computation and memory loads of self-attention operations, transformers are usually employed at downsampled feature resolutions [12-15], which prevents them from processing the subtle textural information and capturing fine-grained long-range dependence at the full image resolution. This limitation is catastrophic for deformable medical image registration, as the full-resolution textural information is crucial to identify subtle anatomy in medical images and find precise pixel-wise spatial correspondence between the anatomical structures. To compensate for this limitation, convolutional layers were employed at the full/half image resolutions in state-of-the-art hybrid CNN-transformer registration networks [12, 14, 15, 21, 22]. Unfortunately, convolutional layers have difficulties in capturing the fine-grained long-range dependence at high resolutions, which results in sub-optimal registration performance.

The desire for modeling long-range dependence has also motivated another research trend toward models based on Multi-layer Perceptrons (MLPs) [23]. By removing self-attention, MLPs are more computationally efficient than transformers while also being able to capture long-range dependence [24-28]. It has been demonstrated that MLPs can be used at the full image resolution to capture fine-grained long-range dependence and have achieved state-of-the-art performance in natural image processing tasks [29] and medical dense prediction tasks [30]. This hints at the potential of leveraging MLPs, as promising alternatives to transformers, to improve deformable image registration. However, MLPs have not been extensively explored for registration and lack the consideration of inductive bias that is crucial for medical registration tasks. For example, existing MLP-based models tend to globally mix feature information along the spatial and channel axes and do not explicitly model the local correlations between features, while the modeling of local correlations has been shown beneficial for deformable registration [31-33]. In addition, state-of-the-art deep registration methods tend to address difficult large deformations via multiple steps of coarse-to-fine registration [9, 11, 14, 22, 32-36, 59], which also has not been investigated by existing MLP-based models.

In this study, we propose a correlation-aware coarse-to-fine MLP-based network (CorrMLP) for deformable image registration. To the best of our knowledge, this is the first study that introduces MLPs for coarse-to-fine deformable registration. In the CorrMLP, we propose a correlation-aware multi-window MLP (CMW-MLP) block, which calculates the local correlations between feature maps and then captures correlation-aware multi-range dependence via multi-window MLP operations. The CMW-MLP block is used in a novel coarse-to-fine registration architecture, where the correlations between images (image-level) and between registration steps (step-level) are both leveraged to realize a correlation-aware coarse-to-fine registration process. Our main contributions are summarized as follows:

- We investigate optimal strategy to leverage MLPs for deformable medical image registration and propose the CorrMLP, to the best of our knowledge, which is the first MLP-based coarse-to-fine registration network.
- We propose the CMW-MLP block, an MLP block that is specifically optimized for deformable registration to capture correlation-aware multi-range dependence.
- We propose a novel correlation-aware coarse-to-fine registration architecture that considers both image-level and step-level correlations to provide enriched contextual information to guide each registration step.

Extensive experiments on two well-benchmarked medical registration tasks (3D inter-patient brain image registration and 4D intra-patient cardiac image registration) with seven public datasets demonstrate that our CorrMLP outperforms state-of-the-art deformable registration methods.

## 2. Related Work
### 2.1. Deformable Medical Image Registration

Early deep registration methods train networks in a fully supervised manner and need ground truth transformations as labels [37, 38]. However, pixel-wise ground truth labels are hard to obtain. To remove the reliance on labels, recent deep registration methods tend to employ image similarity metrics (e.g., mean square error) to train networks in a fully unsupervised manner [7-16]. As one of the most commonly benchmarked registration methods, Balakrishnan et al. [7] proposed a CNN-based network, VoxelMorph, using a hierarchical encoder-decoder architecture similar to Unet [39]. Subsequent studies followed this encoder-decoder architecture [8, 10] and introduced transformers into the networks [12, 13, 21]. Chen et al. [12] proposed a hybrid CNN-transformer registration network, TransMorph, that employs Swin transformer blocks in the encoder. Zhu et al. [13] also proposed a pure transformer-based registration network, Swin-VoxelMorph, that employs a pure Swin transformer architecture similar to Swin-Unet [40]. These studies demonstrate the benefits of modeling long-range dependence for image registration. However, transformers are computationally expensive and were used after 4×4×4 patch embedding. To restore image detail information, TransMorph employed convolutional layers at the full/half image resolutions. Unfortunately, fine-grained long-range dependence still cannot be captured.

Besides the Unet-style direct registration architecture, coarse-to-fine registration architectures were also widely adopted to improve deformable image registration, where multiple steps of registration are performed in a coarse-to-fine manner [9, 11, 14, 22, 32-36]. Mok et al. [9] proposed a laplacian pyramid image registration network (LapIRN), where multiple laplacian pyramid networks were cascaded to perform multiple registration steps. Shu et al. [35] also proposed a ULAE-net to iteratively perform coarse-to-fine registration by running the network for multiple iterations. Recently, Meng et al. [11] proposed a non-iterative coarse-to-fine registration network (NICE-Net) that employs a pyramid network to perform coarse-to-fine registration in a single iteration. The NICE-Net has also been extended to a transformer-based variant, NICE-Trans [22]. This NICE-Trans outperformed its CNN predecessors (NICE-Net) and attained state-of-the-art performance but still did not model fine-grained long-range dependence at full resolution.

In addition, modeling the local correlations between image features has also been shown to facilitate deformable image registration [31-33]. For example, Kang et al. [32] proposed a dual-stream pyramid network (Dual-PRNet++) that enhances deformable image registration by modeling the local 3D correlations between image feature pyramids. However, existing registration methods only considered the correlations between images, while the correlations between registration steps have not been investigated.

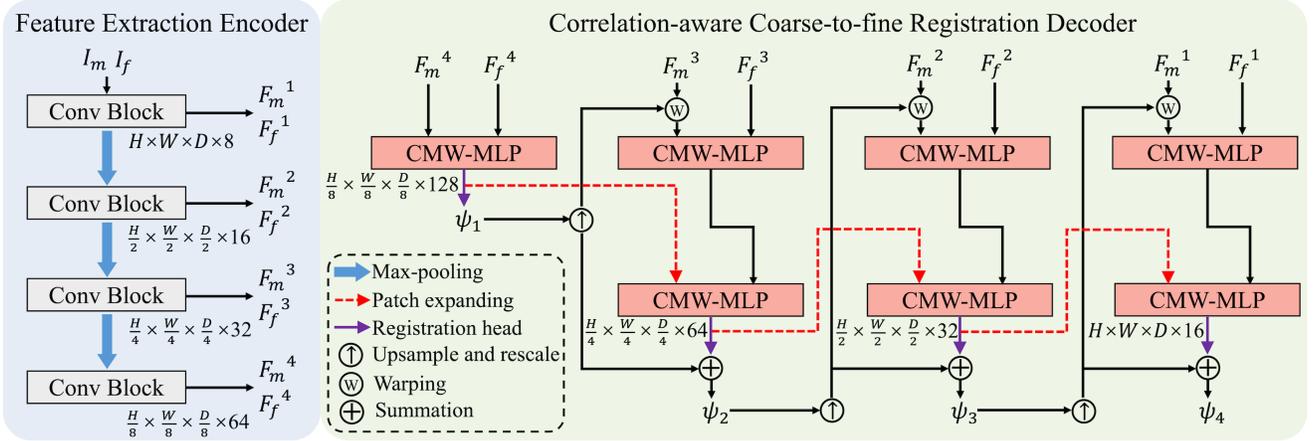

Figure 1: The overall architecture of our CorrMLP. It consists of a CNN-based hierarchical feature extraction encoder and a correlation-aware coarse-to-fine registration decoder based on CMW-MLP blocks.

## 2.2. Multi-layer Perceptrons (MLPs)

MLP-based models have attracted wide attention in the vision community for their capability to capture long-range dependence without relying on self-attention [23]. Early MLPs-based models followed the earlier ViT to process single-scale features after 16×16 patch embedding [24-26]. Tolstikhin et al. [24] proposed an MLP-Mixer that employs MLPs to separately mix channel and spatial information via matrix transpose. Liu et al. [25] also proposed gMLP that introduces spatial gating units for spatial projections. Recently, MLPs have been used in hierarchical pyramid structures [27-30], which extends the application of MLPs to dense prediction. Tu et al. [29] proposed a hierarchical MLP-based model (MAXIM) for low-level natural image processing such as denoising and deblurring. This model achieved state-of-the-art performance on image processing tasks and also demonstrates the feasibility of leveraging MLPs at the full image resolution to capture fine-grained long-range dependence. However, these hierarchical MLP-based models were not optimized for image registration, thus lacking the consideration of inductive bias crucial for deformable medical registration tasks.

So far, MLPs have not been extensively investigated for deformable registration. To the best of our knowledge, only Wang et al. [41] conducted a preliminary study, where both MLP- and transformer-based models were adopted for 2D echocardiography registration. They adopted early MLP-Mixer to perform single-scale image registration and then combined three sub-networks at the 1/4, 1/8, and 1/16 image resolutions for final registration. This method did not employ the recent hierarchical MLP-based models and also cannot capture fine-grained long-range dependence at full/half resolution. Therefore, Wang et al. [41]'s method only achieved similar performance to early VoxelMorph and did not fully reveal the potential of MLPs for deformable medical image registration.

## 3. Method

Image registration aims to find a spatial transformation $\psi$ that warps a moving image $I_m$ to a fixed image $I_f$, so that the warped image $I_{m \circ \psi} = I_m \circ \psi$ is spatially aligned with the fixed image $I_f$. The $I_m$ and $I_f$ are two volumes defined over a $n$-D spatial domain $\Omega \subset \mathbb{R}^n$. In this study, we focus on 3D image registration (i.e., $n = 3$) and assume that the $I_m$ and $I_f$ are uni-modal, single-channel, grayscale images, which is consistent with common deformable medical image registration studies [7-16, 21, 22, 31-36].

We parametrize the deformable registration problem as a function $\mathcal{R}_\theta(I_f, I_m) = \psi$ using CorrMLP (detailed in Section 3.1 and Section 3.2). The $\psi$ is parameterized as a displacement field. The learnable parameters $\theta$ are learned via unsupervised learning (detailed in Section 3.3).

### 3.1. CorrMLP

Figure 1 shows the architecture of our CorrMLP, which consists of a hierarchical feature extraction encoder and a correlation-aware coarse-to-fine registration decoder. The encoder extracts multi-scale hierarchical features from $I_m$ and $I_f$ separately, which is composed of four successive convolutional (Conv) modules with 2×2×2 max-pooling applied between adjacent Conv modules. This feature extraction encoder produces two four-level hierarchical feature pyramids $F_m \in \{F_m^1, F_m^2, F_m^3, F_m^4\}$ and $F_f \in \{F_f^1, F_f^2, F_f^3, F_f^4\}$ from $I_m$ and $I_f$, where the $F_f^i$ and $F_m^i$ are the output of the $i^{th}$ Conv module. Each Conv module contains two 3×3×3 convolutional layers followed by LeakyReLU activation with a parameter of 0.2 and instance normalization. This encoder extracts separate features of $I_m$ and $I_f$, which are then used for multiple coarse-to-fine registration steps.

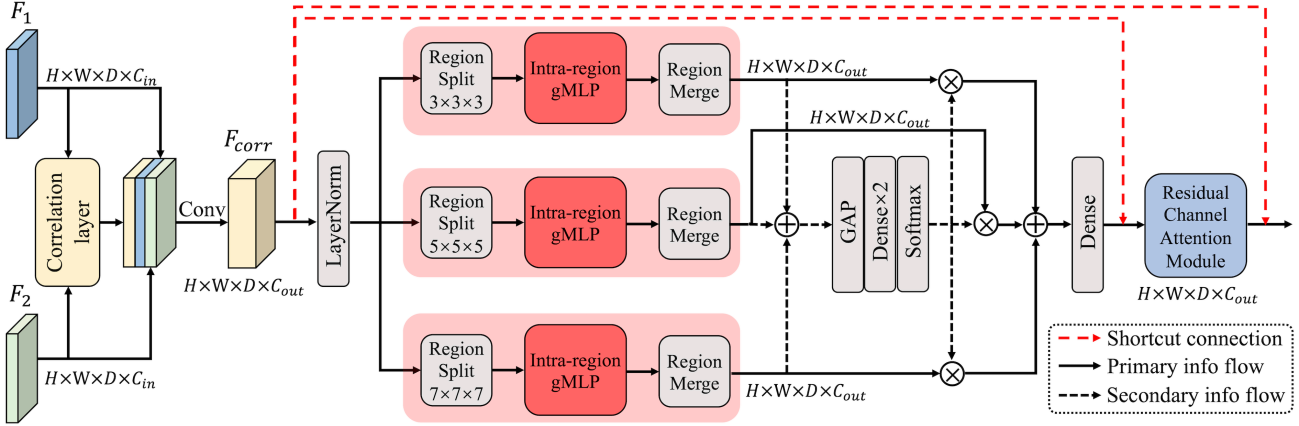

Figure 2: The architecture of our CMW-MLP block. It contains a 3D correlation layer to calculate the local correlations, a multi-window MLP module to capture multi-range dependence, and a residual channel attention module to highlight crucial feature channels.

The decoder leverages the extracted feature pyramids to perform four steps of coarse-to-fine registration using the proposed correlation-aware multi-window MLP (CMW-MLP) blocks (detailed in Section 3.2). In the first step, the $F_m^4$ and $F_f^4$ are fed into a CMW-MLP block to explore their spatial correspondence. The resultant features are fed into a deformable registration head [22] that maps the features into an initial displacement field $\psi_1$. Beginning from the second registration step, there are two CMW-MLP blocks used at each step to model image-level and step-level correlations respectively. Specifically, the $\psi_1$ is used to guide the second registration step, where the warped features $F_m^3 \circ \psi_1$ and $F_f^3$ are fed into the first CMW-MLP block. The second CMW-MLP block then processes the output features from its former CMW-MLP block and the upsampled features derived before the first registration head. The resultant features from the second CMW-MLP block are fed into a deformable registration head. A residual displacement field is produced by the registration head, which is then added to the upsampled $\psi_1$ to form the displacement field $\psi_2$. The above process is repeated for two more times to derive the displacement fields $\psi_3$ and $\psi_4$. The $\psi_4$ is the final registration result $\psi$ that warps the $I_m$ to spatially align with the $I_f$.

The registration decoder employs a novel correlation-aware coarse-to-fine registration architecture, which can perform multiple steps of coarse-to-fine registration within a single iteration of the decoder. Compared with previous coarse-to-fine registration architectures [9, 11, 22], our architecture leverages the correlations between images and between registration steps as supplementary information. The correlations between features of two images have been proven to be effective guidance in establishing the spatial correspondence between the two images [31-33], while the correlations between the steps of coarse-to-fine registration have not been investigated. Our study incorporates step-level correlations to offer important contextual information to each registration step, enabling the model to perform each current registration step based on the awareness of what has been done by its previous steps.

### 3.2. Correlation-aware Multi-window MLP block

Figure 2 illustrates the architecture of our CMW-MLP blocks. It takes two sets of feature maps (denoted by $F_1$ and $F_2$) as input and then explores the potential correspondence between them, which is purposely optimized to capture correlation-aware multi-range dependence for deformable medical image registration. To achieve this, each CMW-MLP block has a 3D correlation layer to calculate the local correlations, followed by a multi-window MLP module to capture correlation-aware multi-range dependence.

Specifically, the $F_1$ and $F_2$ are fed into a 3D correlation layer [32] to obtain a 3D correlation map $C_F$. We set the max displacement $d = 3$ to calculate the local correlations around the 3D neighborhood of 3×3×3. The resultant $C_F$ has the same shape as the $F_1$ and $F_2$ with a channel number of $d^3$=27. The $F_1$, $F_2$, and $C_F$ are concatenated and then fused as a correlation-aware feature map $F_{corr}$ by a 3×3×3 convolutional layer. The $F_{corr}$ is further processed by a multi-window MLP module, where $N$ window-based MLP branches with different window sizes are used in parallel to capture multi-range dependence. In each window-based MLP branch, the feature map is split into non-overlapped regions according to the window size, and then gMLP [25] is adopted to process the features within each region. The outputs of the $N$ MLP branches are fused by channel-wise weighted summation, where the weights are dynamically adjusted via global average pooling (GAP), two-layer MLP, and softmax function. Finally, we followed [29] to employ a residual channel attention module to highlight crucial feature channels, which consists of layer normalization, convolutional layers, LeakyReLU activation, and squeeze-and-excitation (SE) channel attention [42], with a residual connection. In this study, we empirically set $N = 3$ with window sizes of 3×3×3, 5×5×5, and 7×7×7.

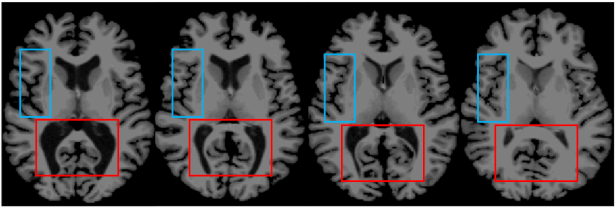

Figure 3: Illustration of local non-linear deformations in medical images. The 2D slices of 3D brain MRI images are presented as examples. The red and blue boxes highlight the regions with relatively large and small deformations among images.

Our CMW-MLP block is purposely optimized with the consideration of crucial prior knowledge for deformable medical image registration. First, deformable registration aims to resolve local non-linear deformations, where the paired pixels usually appear within the local neighborhood of each other [43]. Second, there might exist a wide range of deformations among medical images. As exemplified in Figure 3, the brain Magnetic Resonance Imaging (MRI) images carry both large (red box) and small (blue box) local non-linear deformations among images. Our CMW-MLP block captures multi-range dependence via multiple local window-based MLP branches, enabling it to handle both large and small local deformations.

### 3.3. Unsupervised Learning

The learnable parameters $\theta$ of CorrMLP are optimized using an unsupervised loss $\mathcal{L}$ that does not require labels. The $\mathcal{L}$ consists of two terms $\mathcal{L}_{sim}$ and $\mathcal{L}_{reg}$, where the $\mathcal{L}_{sim}$ is an image similarity term penalizing the differences between the warped image $I_{m \circ \psi}$ and the fixed image $I_f$, while the $\mathcal{L}_{reg}$ is a regularization term encouraging smooth and physically realistic spatial transformations $\psi$.

For the $\mathcal{L}_{sim}$, we adopt negative local normalized cross-correlation (NCC), a similarity metric that has been widely adopted in deformable image registration methods [7-14]. Specifically, let $\hat{I}(\boldsymbol{p})$ denote the local mean intensity of image $I$ in the location $\boldsymbol{p}$:

$$\hat{I}(\boldsymbol{p}) = \frac{1}{n^3}\sum_{\boldsymbol{p}_i} I(\boldsymbol{p}_i), \quad (1)$$

where the $\boldsymbol{p}_i$ iterates over a $n^3$ neighboring region around $\boldsymbol{p}$, with $n = 9$ in our experiments. Then, the $\mathcal{L}_{sim}$ between the $I_f$ and $I_{m \circ \psi}$ is defined as:

$$\mathcal{L}_{sim}(I_f, I_{m \circ \psi}) =$$
$$-\sum_{\boldsymbol{p} \in \Omega} \frac{\left(\sum_{\boldsymbol{p}_i}[I_f(\boldsymbol{p}_i)-\hat{I}_f(\boldsymbol{p})][I_{m \circ \psi}(\boldsymbol{p}_i)-\hat{I}_{m \circ \psi}(\boldsymbol{p})]\right)^2}{\left(\sum_{\boldsymbol{p}_i}[I_f(\boldsymbol{p}_i)-\hat{I}_f(\boldsymbol{p})]^2\right)\left(\sum_{\boldsymbol{p}_i}[I_{m \circ \psi}(\boldsymbol{p}_i)-\hat{I}_{m \circ \psi}(\boldsymbol{p})]^2\right)}. \quad (2)$$

For the $\mathcal{L}_{reg}$, we impose a diffusion regularizer on the $\psi$ to encourage its smoothness:

$$\mathcal{L}_{reg}(\psi) = \sum_{\boldsymbol{p} \in \Omega} ||\nabla \psi(\boldsymbol{p})||^2, \quad (3)$$

where the $\nabla$ is the spatial gradient operator.

The final unsupervised loss $\mathcal{L}$ is defined as:

$$\mathcal{L}(I_f, I_m, \psi) = \mathcal{L}_{sim}(I_f, I_{m \circ \psi}) + \lambda \mathcal{L}_{reg}(\psi), \quad (4)$$

where the $\lambda$ is a regularization parameter balancing the registration accuracy and transformation smoothness.

## 4. Experimental Setup

### 4.1. Datasets and Preprocessing

We evaluated our CorrMLP with two well-benchmarked deformable image registration tasks (3D inter-patient brain image registration and 4D intra-patient cardiac image registration), involving seven public medical datasets:

For brain image registration, we adopted six public 3D brain MRI datasets that have been widely used to evaluate medical image registration [7-16]. We followed the dataset settings in [11]: a total of 2,656 brain MRI images acquired from four public datasets (ADNI [44], ABIDE [45], ADHD [46], and IXI [47]) were used for training; two public brain MRI datasets with anatomical segmentation (Mindboggle [48] and Buckner [49]) were adopted for validation and testing. The Mindboggle dataset contains 100 MRI images and was randomly split for validation/testing with a ratio of 50%/50%. The Buckner dataset contains 40 MRI images and was used for independent testing. We followed the existing studies [7-15] to perform inter-patient registration for evaluation, where 100 image pairs were randomly picked from each of the Mindboggle and Buckner testing sets, resulting in 200 testing image pairs in total. We performed standard brain MRI preprocessing procedures, including brain extraction, intensity normalization, and affine registration by FreeSurfer [49] and FLIRT [50]. All images were affine-transformed and resampled to align with the MNI-152 brain template [51] with 1mm$^3$ isotropic voxels, which were then cropped into 144×192×160.

For cardiac image registration, we adopted the public ACDC dataset [52] that contains 4D cardiac cine-MRI images of 150 patients. Each 4D cine-MRI image contains tens of 3D frames acquired from different time-points, including End-Diastole (ED) and End-Systole (ES) frames with segmentation labels of the left ventricular cavity, right ventricular cavity, and myocardium. The ED is defined as the first frame when the mitral valve closes and the ES is defined as the first frame when the aortic valve closes. ED and ES delineate the two ends of a cardiac cycle and show the largest deformation in the cardiac cycle [53]. The ACDC dataset provides 100 cine-MRI images in the training set and 50 cine-MRI images in the testing set, where we randomly divided the training set into 90 and 10 cine-MRI images for training and validation and used the provided testing set for testing. Following [54, 55], we aim to register the ED and ES frames of the same patient. The intra-patient ED and ES frames were registered with each other (ED-to-ES and ES-to-ED), resulting in 100 testing image pairs derived from the testing set. All cine-MRI frames were resampled with a voxel spacing of 1.5×1.5 ×3.15mm$^3$ and cropped to 128×128×32 voxels around the center. The voxel intensity was normalized to range [0, 1] through max-min normalization.

### 4.2. Implementation Details

Our CorrMLP was implemented using PyTorch on an NVIDIA GeForce RTX 4090 GPU with 24 GB memory. We adopted an ADAM optimizer with a learning rate of 0.0001 and a batch size of 1. The regularization parameter $\lambda$ was set as 1. For brain image registration, the CorrMLP was trained for 100,000 iterations with inter-patient image pairs randomly picked from the training set. For cardiac image registration, the CorrMLP was trained for 40,000 iterations with intra-patient image pairs that consist of two frames randomly picked from the same cine-MRI image. At the last 10,000 iterations, the CorrMLP was trained with intra-patient image pairs consisting of only ED and ES frames, which optimizes the model to register ED and ES frames. We performed validation after every 1,000 training iterations and preserved the model achieving the highest validation result for final testing. Our code is available at *https://github.com/MungoMeng/Registration-CorrMLP*.

### 4.3. Comparison Methods

Our CorrMLP was extensively compared to state-of-the-art deformable image registration methods, including two traditional optimization-based registration methods and ten deep registration methods. The two traditional methods are SyN [17] and NiftyReg [18], and we ran them using cross-correlation as the similarity measure. The included deep registration methods are VoxelMorph [7], TransMorph [12], Swin-VoxelMorph [13], TransMatch [15], LapIRN [9], ULAE-net [35], Dual-PRNet++ [32], NICE-Net [11], NICE-Trans [22], and SDHNet [36]. Moreover, we also incorporated MAXIM [29], a state-of-the-art MLP-based natural image processing method, into the comparison. We adapted it to deformable registration by reimplementing it as a 3D model and modifying its output layer to produce displacement fields. Due to its highly complex architecture requiring large GPU memory, the MAXIM is difficult to apply to 3D brain MRI images and we, therefore, evaluated it only on cardiac image registration with relatively smaller image size. All deep learning methods were trained with the same loss functions and data split settings as ours.

### 4.4. Evaluation Metrics

We adopted standard evaluation metrics commonly used in medical image registration studies [7-16] and in related registration challenges [56, 57]. The registration accuracy was evaluated using the Dice similarity coefficients (DSC) between the segmentation labels of the fixed and warped images ($I_f$ and $I_{m \circ \psi}$). A two-sided $P$ value less than 0.05 is considered to indicate a statistically significant difference between two DSCs. The smoothness and invertibility of the predicted spatial transformations were evaluated using the percentage of Negative Jacobian Determinants (NJD) [58]. There often is a trade-off between registration accuracy and transformation smoothness by adjusting the regularization parameter [8, 11]. Therefore, registration methods should be evaluated in terms of both DSC and NJD.

### 4.5. Experimental Designs

Our CorrMLP was compared to the existing deformable registration methods for both brain and cardiac image registration. Then, we performed two ablation studies to analyze the individual contributions of our coarse-to-fine registration architecture and CMW-MLP block.

In the first ablation study, we explored the effects of different architecture designs on registration performance. We built a Unet-like MLP-based registration network as the baseline and followed the popular VoxelMorph and TransMorph to name it as MLPMorph. The MLPMorph adopted an Unet-like encoder-decoder architecture similar to TransMorph, which uses MLP blocks in the encoder and CNN blocks in the decoder. Multi-window MLP (MW-MLP) blocks (a variant of our CMW-MLP that removes the 3D correlation layer) were used in the MLPMorph to extract hierarchical multi-scale image features beginning from the full image resolution. We also used MW-MLP blocks in our coarse-to-fine registration architecture, resulting in three degraded models that cannot leverage the image-level and/or step-level correlation information.

In the second ablation study, we studied the registration performance when using different MLP blocks. We kept our network architecture and replaced our CMW-MLP blocks with existing MLP blocks including Spatial-shift MLP (S$^2$-MLP) [26], Sparse MLP (sMLP) [27], Hire-MLP [28], Swin-MLP [20], and the multi-axis gated MLP used in MAXIM [29]. We also attempted to remove some of the MLP branches in our CMW-MLP blocks to validate the effectiveness of the multi-window MLP design.

## 5. Results and Discussion

### 5.1. Comparison with Existing Methods

Table 1 and Table 2 show the quantitative comparison between the CorrMLP and existing registration methods for brain and cardiac image registration. Among the existing methods, transformer-based TransMorph and Swin-VoxelMorph achieved higher DSCs than the widely benchmarked CNN-based VoxelMorph, which validates the benefits of capturing long-range dependence for image registration. Furthermore, the single-stage and multi-stage (×3) variants of MAXIM were evaluated for cardiac image registration, and both of them achieved higher DSCs than the transformer-based direct registration methods (Table 2). This illustrates the superiority of MLP-based models for deformable registration tasks. However, the MAXIM is particularly optimized for natural image processing tasks, which is not competitive with state-of-the-art coarse-to-fine registration methods due to the lack of inductive bias crucial for medical image registration.

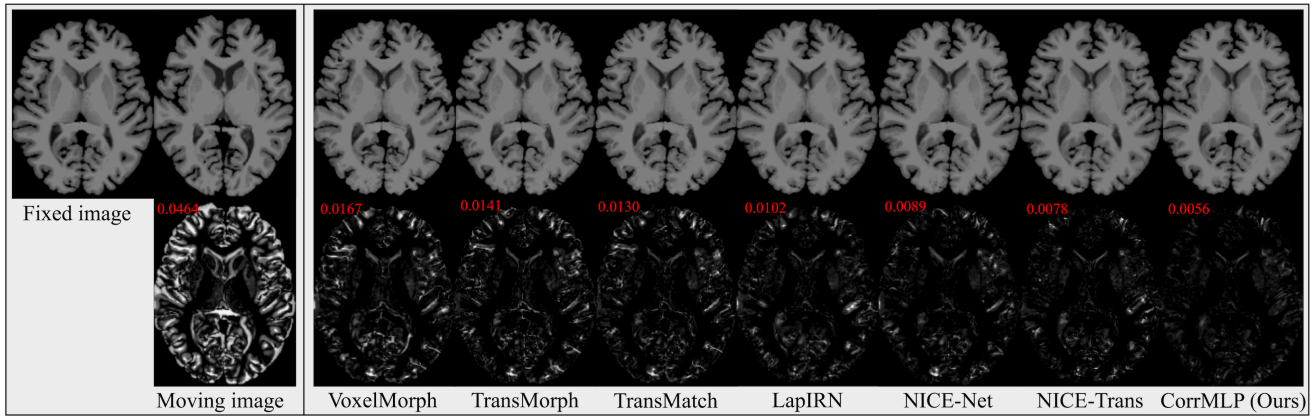

Figure 4: Qualitative comparison for brain image registration. Below each image is an error map that shows the intensity differences from the fixed image, with the mean absolute error placed in the upper left corner. A cleaner error map indicates a better registration result.

| Method | | Mindboggle dataset | | Buckner dataset | | Runtime | |
|---|---|---|---|---|---|---|---|
| | | DSC ↑ | NJD (%) ↓ | DSC ↑ | NJD (%) ↓ | CPU (s) | GPU (s) |
| Before registration | | 0.347* | / | 0.406* | / | / | / |
| SyN [17] | Traditional | 0.534* | 1.956 | 0.567* | 1.874 | 3427 | / |
| NiftyReg [18] | Traditional | 0.569* | 2.364 | 0.611* | 2.175 | 159 | / |
| VoxelMorph [7] | CNN, direct | 0.552* | 2.532 | 0.589* | 2.220 | 2.84 | 0.23 |
| Swin-VoxelMorph [13] | Transformer, direct | 0.566* | 2.254 | 0.605* | 2.016 | 5.67 | 0.52 |
| TransMorph [12] | Transformer, direct | 0.571* | 2.400 | 0.608* | 2.183 | 3.68 | 0.35 |
| TransMatch [15] | Transformer, direct | 0.578* | 2.036 | 0.622* | 1.995 | 3.06 | 0.28 |
| LapIRN [9] | CNN, coarse-to-fine | 0.605* | 2.164 | 0.632* | 2.112 | 4.97 | 0.46 |
| ULAE-net [35] | CNN, coarse-to-fine | 0.610* | 2.000 | 0.640* | 1.940 | 5.37 | 0.51 |
| Dual-PRNet++ [32] | CNN, coarse-to-fine | 0.608* | 2.424 | 0.636* | 2.195 | 4.61 | 0.44 |
| SDHNet [36] | CNN, coarse-to-fine | 0.598* | 1.872 | 0.634* | 1.843 | 3.24 | 0.26 |
| NICE-Net [11] | CNN, coarse-to-fine | 0.618* | 2.043 | 0.643* | 1.963 | 3.55 | 0.32 |
| NICE-Trans [22] | Transformer, coarse-to-fine | 0.625* | 2.324 | 0.649* | 2.277 | 4.02 | 0.37 |
| CorrMLP (Ours) | MLP, coarse-to-fine | **0.642** | **1.821** | **0.661** | **1.788** | 5.48 | 0.49 |

Table 1: Quantitative comparison for brain image registration. The best results in each dataset are in bold. ↑: the higher is better. ↓: the lower is better. *: $P<0.05$, in comparison to CorrMLP.

Moreover, coarse-to-fine registration methods attained consistently higher DSCs than direct registration methods, validating the superiority of coarse-to-fine registration. By employing transformers in a coarse-to-fine registration architecture, the NICE-Trans achieved higher DSCs than all other comparison methods. Nevertheless, our CorrMLP outperformed the NICE-Trans and achieved significantly higher DSCs than all comparison methods. Our CorrMLP obtained similar NJDs to other registration methods, which means that our CorrMLP did not sacrifice transformation smoothness for registration accuracy. The runtime of each method is also reported in Table 1 and Table 2, which shows that our CorrMLP is much faster than the traditional methods (SyN and NiftyReg), and its runtime is similar to the existing deep registration methods, allowing real-time registration with GPUs (<0.5s for one image pair).

| Method | ACDC | | Runtime | |
|---|---|---|---|---|
| | DSC ↑ | NJD (%) ↓ | CPU (s) | GPU (s) |
| Before registration | 0.590* | / | / | / |
| VoxelMorph [7] | 0.754* | 0.440 | 0.36 | 0.02 |
| Swin-VoxelMorph [13] | 0.763* | 0.412 | 0.91 | 0.08 |
| TransMorph [12] | 0.768* | 0.492 | 0.59 | 0.05 |
| TransMatch [15] | 0.770* | 0.425 | 0.55 | 0.04 |
| MAXIM [29] | 0.785* | 0.437 | 1.82 | 0.17 |
| MAXIM×3 [29] | 0.788* | 0.716 | 5.45 | 0.51 |
| LapIRN [9] | 0.790* | 0.454 | 0.77 | 0.06 |
| ULAE-net [35] | 0.792* | 0.447 | 0.86 | 0.07 |
| Dual-PRNet++ [32] | 0.777* | 0.479 | 0.75 | 0.06 |
| SDHNet [36] | 0.789* | 0.395 | 0.45 | 0.03 |
| NICE-Net [11] | 0.785* | 0.443 | 0.49 | 0.04 |
| NICE-Trans [22] | 0.799* | 0.473 | 0.64 | 0.05 |
| CorrMLP (Ours) | **0.810** | **0.389** | 0.83 | 0.07 |

Table 2: Quantitative comparison for cardiac image registration. The best results are in bold. ↑: the higher is better. ↓: the lower is better. *: $P<0.05$, in comparison to CorrMLP.

Figure 4 shows a qualitative comparison of brain image registration. Consistent with the quantitative results (Table 1), the registration result produced by our CorrMLP is the most consistent with the fixed image, thus resulting in the cleanest error map among all the compared methods.

### 5.2. Analysis of Architecture Designs

Table 3 presents the DSC results of the ablation study on architecture designs. The NJD results are omitted as all methods adopted the same regularization settings and achieved similar NJDs. Our baseline MLPMorph has already outperformed VoxelMorph and TransMorph by a large margin. We attribute this to the fact that the TransMorph uses convolutional layers to process full/half-resolution and thus has difficulty in capturing fine-grained long-range dependence at high resolutions, while the MLPMorph employs MLP blocks to directly process high-resolution features. This demonstrates the superiority of MLPs over transformers and CNNs on deformable image registration: MLP blocks can capture fine-grained long-range dependence at high resolutions, which is crucial for finding precise dense correspondence as high-resolution features provide richer detailed information.

By employing MLP blocks in a correlation-aware coarse-to-fine registration architecture, our CorrMLP also outperformed MLPMorph by a large margin. To validate the effects of correlation information on coarse-to-fine registration, we separately removed the image-level and step-level correlation information and found that removing either information degraded the registration performance. This ablation study suggests that both image-level and step-level correlations are beneficial for coarse-to-fine registration and also proves that our correlation-aware coarse-to-fine registration architecture outperforms the existing coarse-to-fine architectures that do not leverage these correlation information.

### 5.3. Analysis of MLP Blocks

Table 4 presents the DSC results of the ablation study on MLP blocks. The NJD results are omitted as all methods adopted the same regularization settings and achieved similar NJDs. Replacing our CMW-MLP blocks with five different existing MLP blocks all resulted in lower DSCs, which demonstrates the effectiveness of our CMW-MLP block. This effectiveness can be partly attributed to the awareness of correlation information in modeling long-range dependence. However, even when the correlation layer was removed, the MW-MLP block also outperformed the five existing MLP blocks, which implies that our multi-window MLP design is also beneficial for deformable image registration. To further validate this, we separately removed one of the MLP branches in CMW-MLP blocks and found that removing any MLP branch degraded the registration performance. We also attempted to add an extra MLP branch with a window size of 9×9×9, but we did not identify any further improvements in performance. This suggests that a 7×7×7 MLP branch has been sufficient to capture large deformations, while the 3×3×3 and 5×5×5 MLP branches are crucial to capture subtle deformations.

| Method | Mindboggle | Buckner | ACDC |
|---|---|---|---|
| VoxelMorph [7] | 0.552 | 0.589 | 0.754 |
| TransMorph [12] | 0.571 | 0.608 | 0.768 |
| MLPMorph (Ours) | 0.604 | 0.632 | 0.780 |
| No correlation | 0.628 | 0.650 | 0.800 |
| Only image-level correlation | 0.637 | 0.657 | 0.806 |
| Only step-level correlation | 0.634 | 0.655 | 0.805 |
| CorrMLP (Ours) | **0.642** | **0.661** | **0.810** |

Table 3: DSC results of the ablation study on architecture designs. The best results are in bold.

| MLP block | Mindboggle | Buckner | ACDC |
|---|---|---|---|
| $S^2$-MLP [26] | 0.621 | 0.644 | 0.794 |
| sMLP [27] | 0.622 | 0.645 | 0.794 |
| Hire-MLP [28] | 0.620 | 0.643 | 0.793 |
| Swin-MLP [20] | 0.624 | 0.646 | 0.797 |
| Multi-axis gated MLP [29] | 0.625 | 0.647 | 0.798 |
| MW-MLP (Ours) | 0.628 | 0.650 | 0.800 |
| No 3×3×3 MLP branch | 0.639 | 0.657 | 0.808 |
| No 5×5×5 MLP branch | 0.635 | 0.654 | 0.805 |
| No 7×7×7 MLP branch | 0.637 | 0.655 | 0.806 |
| CMW-MLP (Ours) | **0.642** | **0.661** | **0.810** |

Table 4: DSC results of the ablation study on MLP blocks. The best results are in bold.

### 6. Conclusion

In this study, we have shown the effectiveness of MLPs for deformable medical image registration by developing the first MLP-based coarse-to-fine registration network (CorrMLP). In the CorrMLP, we introduce a correlation-aware multi-window MLP (CMW-MLP) block and use it in a novel coarse-to-fine registration architecture that takes into account both image-level and step-level correlations. Extensive experiments on both brain and cardiac image registration show that, with the CMW-MLP block and the correlation-aware coarse-to-fine registration architecture, our CorrMLP can outperform state-of-the-art registration methods. Furthermore, we suggest that our CMW-MLP block can serve as a general MLP block applying to various network architectures for image registration tasks, and our CorrMLP also can apply to multi-modal registration tasks such as multi-parametric brain MRI registration [60].

**Acknowledgement:** This work was supported in part by Australian Research Council (ARC) Grant DP200103748.